\documentclass[12pt]{article}

\usepackage[top=3cm, bottom=3cm, left=3cm, right=3cm]{geometry}

\usepackage[T1]{fontenc}
\usepackage{amsthm}
\usepackage{paralist}
\PassOptionsToPackage{hyphens}{url}

\newcommand{\frost}{\textsf{FROST}}
\newtheorem{principle}{Principle}

\newcommand{\url}[1]{{\small {\tt #1}}}

\begin{document}

\begin{center}
{\Large \bf Owner-centric sharing of physical resources, data,
  and data-driven insights in digital ecosystems}
\end{center}

\bigskip
\begin{center}
{\small Kwok Cheung and Laurence Kirk\\
Extropy Ltd., Oxford, UK\\
$\{$kwok, laurence$\}$@extropy.io}
\end{center}

\begin{center}
{\small Michael Huth\\
Imperial College London, London, UK (and XAIN AG)\\
m.huth@imperial.ac.uk}
\end{center}

\begin{center}
{\small Leif-Nissen Lundb\ae k, Rodolphe Marques, and Jan Petsche\\
XAIN AG, Berlin, Germany\\
$\{$leif, rodolphe.marques, jan.petsche$\}$@extropy.io}
\end{center}

\bigskip

\begin{quote}
{\bf Notice:} This is a plain-latex and reduced version of the paper with the same title which appeared in \emph{The 24th ACM Symposium on Access Control Models and Technologies}, June 3--6, 2019, Toronto, ON, Canada. The DOI of that conference paper is {\bf 10.1145/3322431.3326326} where the full URL is {\bf https://dl.acm.org/citation.cfm?doid=3322431.3326326}.
\end{quote}

\bigskip
\begin{abstract}
We are living in an age in which digitization will connect more and
more physical assets with IT systems and where IoT endpoints will
generate a wealth of valuable data. Companies, individual users, and
organizations alike therefore have the need to control their own
physical or non-physical assets and data sources. At the same time,
they recognize the need for, and opportunity to, share access to such
data and digitized physical assets. This paper sets out our technology
vision for such sharing ecosystems, reports initial work in that
direction, identifies challenges for realizing this vision, and seeks
feedback and collaboration from the academic access-control community
in that R\&D space.
\end{abstract}

\bigskip
\begin{quote}
\end{quote}

\newpage{}

\section{Introduction}

There is little doubt that the increased digitization of our physical and social world has already had profound impact~--~economically, socially, and technically. The adoption rate of smart-phones and their increased capabilities and richness of user interactions through apps are just one source of evidence for this claim. Digitization is also changing the manufacturing plants and their processes, vehicle function and interaction with the environment, and general consumer expectations of service provisions. For the latter, progressive web apps are an example of how today's users wish to interact with products and services~--~where the ability of an app to adapt to user-specific interaction history and needs is seen as a competitive advantage.

The increased connectedness that this digitization of our lives brings, and its shift from product purchase to user-centric service consumption drives, in part, the creation of nascent sharing economies. For example, throughout the globe there are a host of providers for e-bikes, e-scooters, and cars operating locally where clients can use these resources on demand for flexible periods of time. And with the growth rate of sales of new cars predicted to fall to about 2 percent globally by 2030~\cite{AutoMcKinsey16}, Original Equipment Manufacturers (OEM) are incentivized to find means of increasing the monetary value of a manufactured vehicle. One way in which an OEM can do this is to not sell the car at all, but rather operate a service by which the car can be used by many, thus increasing the active usage time and the revenue stream generated by usage of the vehicle.

The sharing economy is not limited to the consideration of  Business-To-Consumer (B2C) business models. A decentralized app for accommodation sharing, e.g., may enable Consumer-To-Consumer (C2C) business models for the sharing of rooms, apartments or other facilities such that clients can directly formulate the access conditions to their own data and physical resources without a central organization having already access to such data and resources.\footnote{The governance of this platform would be exercised by some parties that would hold power over running that platform, but that should not enable these parties to circumvent access controls that operate within that platform itself.} Such an approach requires, though, that clients as owners of physical resources can freely delegate the access of their own resources to others~--~under specific conditions that typically include or trigger financial transactions.

Of course, the sharing economy is neither confined to mobility and transportation nor to B2C and C2C business models. A typical Business-To-Business (B2B) case is found in ecosystems in which companies already have established working relationships and work-flows across  organizations. Generally, such an ecosystem lacks the technical means of judiciously sharing access to physical resources, data or  data-driven insights \emph{across}  organizational boundaries in a digital, fine-grained, and (semi)-automated manner.

This gap presents a business opportunity as the provision of such digital sharing processes can reduce cost (e.g. removal of manual data entry on the receiving end), improve security (e.g.~prevention of ad-hoc data sharing such as through email attachments), and create economic value (e.g.~by leasing under-utilized physical resources).

For example, a company that operates train infrastructure may want to share some data sets with a train manufacturer, and share different data sets and access to physical devices deployed on tracks with a service company that maintains these tracks. But even if these parties had already understood the specifics of what should be shared with whom under which circumstances, they would have neither the technical means of operationalizing this understanding, nor of verifying the trustworthiness of such an operationalization.

It is clear that a technical solution to this needs to provide sufficient interoperability. Non-technical issues such as social, regulatory or cognitive ones also may need to be overcome; e.g., company culture may make it hard to contemplate that there is value in sharing data or resources with other, perhaps competing, companies. Regulation may prevent such sharing or put demands on its operationalization and auditing. And understanding what should be shared and exactly how and with whom can create cognitive complexity that needs to be managed~--~regardless of what specification language is being used for access controls.

The engagement with our clients and potential customers has made us realize that there is a definite need for technology that can provide such owner-centric means of sharing physical resources, data, and data-driven insights such that participating parties can agree on the specifics of their sharing topology and its controls. The realization of the vision we offer in this paper faces several challenges, some of which we will highlight subsequently. The intent of this paper is to share this vision and its challenges in order to gain feedback and to encourage collaboration on the needed research and development (R\&D) in order to make this vision a reality.

\paragraph{Outline of Paper} Our vision and its guiding principles are the subject of Section~\ref{section:vision}. Our work done so far towards realizing this vision is discussed in Section~\ref{section:workdonesofar}. In Section~\ref{section:challenges}, we explore some challenges that attaining this vision faces. Section~\ref{section:academic} details research topics in which we think the academic access-control community can make vital contributions for our vision of sharing ecosystems. Representative related work  is summarized in Section~\ref{section:relatedwork} and the paper concludes in Section~\ref{section:conclusion}.

\section{Our Vision and its Principles}
\label{section:vision}

We now discuss our vision by formulating its guiding principles and technical ambition, while also explaining the values that determine these principles. 

First, we observe that there is an apparent tension between centralization of service provision and genuine control of owners over their digital or physical assets. This can be seen, e.g., in the manner in which big tech companies such as Facebook, Apple, Amazon, Netflix, and Google give their users only limited control over data generated by them. 

However, it is the ability to control access and to manage the life cycle of such data that amounts to what most of us would understand to be \emph{data ownership}. In UK law, e.g., there seems to be nothing on the Statute Books that refers to data \emph{ownership} directly but rather to the \emph{control} of data. Indeed, it is unclear whether users, owners or manufacturers (say of a car) could \emph{own} a piece of data from that asset~--~say under current EU law. It is therefore conceivable that Internet of Things (IoT) data may have multiple owners. Moreover, future owners may neither be human subjects nor legal entities~--~e.g.\ autonomous machines.

Our first principle assumes that such legal clarity of ownership and data creation will be established in the future:
\begin{principle}[Owner-Centric Data Handling]
  The creators of data should be in genuine control over their data.
\end{principle}

Note that this principle does not rule out that data owners may give away their own data or even control over it~--~be it for free as a social good, for money or for some other motive. But it does mean that the owners of such data, e.g.~of sensor data from a vehicle, are in full control of that data as soon as it gets created~--~and that these owners can decide to delegate to others access of such data. We emphasize that this applies not only to owners as individuals but equally when the owner is a legal entity such as a limited company that hosts its data in a cloud environment but on premise.

An owner-centric approach also places great importance onto data privacy: data controls should default to using access-control policies that protect the user, and where deviations from defaults are at the discretion of data owners.

For \emph{physical} assets, it is often said that ``possession is nine-tenths of the law''. We believe that this expression will need to be revised in the digital future, since most physical assets of value will have micro-controllers embedded in them that guard access to those assets and their functions based on digital credentials. This leads us to the following principle:

\begin{principle}[Owner-Centric Access to Physical Assets]
  The owners of physical assets should be in control over these assets and over the ability to delegate such access controls.
\end{principle}

This principle seems like a truism. But it should also apply when the owner is not in physical proximity of the asset and when the owner is a legal entity (e.g.~as might be the case for a car that ``owns itself''). The delegation mechanism stated in this principle gives rise to social and technical complexity, manifesting itself in two types: delegation of
\begin{compactitem}
  \item basic access rights such as ``my daughter may use my Porsche on weekends before 8pm'',
  \item specifying entire access-control policies, e.g.~when the OEM would say that ``this car dealer may install and run an access-control policy on that vehicle as well".  
\end{compactitem}

Transfer of ownership may be seen as a special case of that second type of delegation. In the future, the OEM will certainly want to maintain a numerical passport of the vehicle it manufactured, e.g.~in order to do predictive maintenance and to improve future versions of that particular car model. But such access to data needs to be consistent with whatever policy the car dealer, e.g.~as legal owner of the car, would want to put onto the vehicle. Similarly, the policy of a leaseholder (to whom the car dealer may have delegated the right to install such a policy on the vehicle) would need to be consistent with that of the car dealer and with the one of the OEM. This leads us to another principle:

\begin{principle}[Programmable Sharing Ecosystem]
  Technology should enable the expression of delegation and access relationships that are consistent with each other and reflect the needs of owners and consumers of data and other assets. 
\end{principle}

Let us explore this principle in a case in which two companies (X and Y) want to share access to data sets about driving behavior over their own vehicle models. They have an incentive to share the knowledge that resides within such data sets, e.g.~in order to improve the quality of their machine-learning algorithms for semi-automatic driving. 

Yet companies X and Y will never agree to actually share these data sets as in a ``may read/may copy'' modality that would move such data onto the other company's premises. But they may agree to running a machine-learning algorithm on their own data sets on behalf of the other company and to then report back the learned mathematical model. The principle says that the technology should allow these companies to leave their data sets on premise, to run machine-learning algorithms from the other company on their own data sets, and to have a fair way of reporting those results back. 

Fairness may amount to selling the model for an adequate price, to using an escrow service if both companies expect to receive models from the other company's data sets or to relying on some other mutually agreed upon trust mechanism. We believe that the \emph{``may train''} modality for this use case of data access~--~which is weaker than ``write'' and ``read''~--~ is of independent interest for the access-control community.

This principle also applies when actors share a physical resource, as our use case of the OEM, the car dealer, and the leaseholder of a vehicle illustrated. In that use case, the principle assures that these stakeholders have the technological means of mapping their social or contractual understanding of access to vehicle data and function consistently onto the vehicle technology and its communication interfaces. 

That use case also demonstrates that more  data will be generated from IoT endpoints, and more computation will be pushed to IoT edges. This technology trend, and general security considerations, lead us to our next principle:

\begin{principle}[Physical Proximity of Access-Control]
  Access-control, including the verification of both delegation and change management of controls, should be enforced close to or embedded into the assets that are subject to these controls.
\end{principle}

Vehicles would thus compute and enforce access-control decisions within their internal vehicle systems, potentially also with offline capabilities for communicating and verifying policy state. And enterprises may host a server that provides similar functionality in controlling access to local data sets and the reporting of data-driven insights to external parties.

From a business perspective, there is an incentive to realize such access controls with the same  software for the micro-controller and the server environments; this would decrease development costs, would ease maintenance, and so forth. We therefore formulate a principle rooted in considerations of code assurance and economics:

\begin{principle}[Software is Executable on IoT and ICT]
  Software that can reliably realize such access-control for both micro-controllers and servers has a competitive advantage.
\end{principle}

The importance of this principle should not be underestimated; e.g., it rules out the use of eXtensible Markup Language (XML) technology for building such software, since XML is not supported or too demanding in most micro-controller environments. It also suggests that a balance needs to be struck between the expressiveness of programming languages running on servers and the restrictions in terms of storage and available technology stacks and toolings found on commercial micro-controllers.

The access-control software we envision serves as an abstraction layer on top of existing controls, e.g.\ those within a vehicle network or in an enterprise network. Therefore, this principle is consistent with the use of server-side technology such as eXtensible Access Control Markup Language (XACML) at lower layers of abstraction.

The use cases we described so far, Company X/Company Y and OEM, Car Dealer, and Leaseholder, may be interpreted as closed systems in and of themselves. We prefer to call them \emph{biotopes} that may choose to connect with other such sharing biotopes within larger sharing ecosystems, e.g.~emerging data marketplaces. From a technology perspective, this means that we also want to accommodate \emph{open} sharing systems. For example, a company that manufactures batteries may want to share battery usage data with a provider of electric charging stations for e-scooters, and e-scooter users may share their location data with advertisers for free battery charging~--~should they deem this deal to be acceptable.

\begin{principle}[Open Sharing Ecosystems]
  Sharing systems can connect to other sharing systems in an open network with trustworthy consensus mechanism.
\end{principle}

We believe that decentralized consensus architectures as found in Distributed Ledger Technology (DLT) have good potential in this regard: important system state needs to have consensus in order to make the system function reliably (under faults) and resiliently (under attacks from active adversaries). Known consensus algorithms provide such reliability and resiliency at high cost: they may not scale to desired network or transaction size (e.g.~Byzantine algorithms such as PBFT~\cite{DBLP:conf/osdi/CastroL99}), they may be proprietary (e.g.~Hashgraph at \url{https://hedera.com}) or they may be wasting energy (e.g.~Proof of Work~\cite{DBLP:conf/crypto/DworkN92,nakamoto08}). This leads us to another principle:

\begin{principle}[Sustainable Technology]
  Sharing ecosystems should use technology that makes these systems sustainable.
\end{principle}

For example, we do not believe that consuming energy at the scale of a major industrial economy is a good way of implementing consensus in an open network for sharing ecosystems, as seems to be the case for Proof of Work in the Bitcoin cryptocurrency~\cite{nakamoto08}. And we also do not accept that the incentive mechanisms that such open systems typically require should lead to unfair treatment of some network participants over others, e.g.~as seen in the dynamics of mining for cryptocurrencies based on Proof of Work~--~where it is no longer feasible to mine as a private citizen with standard computing equipment. This motivates our next principle:

\begin{principle}[Fair and Democratic Sharing in Open Networks]
  Technology should enable a democratic and fair participation in open sharing ecosystems.
\end{principle}

This principle is one based on values that are of a political or civic nature. It does not mean that network participants would all be equal, but it does imply that such networks should have transparent and fair governance structures in which network participants may gain influence based on the good contributions they have made to that network over time. For example, network participants may gain influence by contributing to the creation of consensus in the network state. However, the principle is also consistent with an open network that has primarily a commercial purpose and whose governance and membership models were co-designed with initial commercial stakeholders.

Unlocking the economic value of such networks will certainly require  sufficient interoperability between IoT and ICT environments that allows for the flow of data and data-driven insights as well as for the evaluation and enforcement of access-control policies that regulate such flow~\cite{IoT15,Nicolescu2018}.

The lack of interoperability in IoT is a well recognized problem. Standardization efforts, through national and international bodies, are under way in order to address this issue (see e.g.~the International Organization for Standardization and its Technical Committee~307 on blockchain and distributed ledger technologies). While we applaud such initiatives, we also recognize that it is difficult (politically and technologically) to create a globally recognized official standard that provides such interoperability all the way from IoT edges to ICT servers and storage environments. 

Rather, we deem it to be more realistic to build lightweight software development kits (SDK) that can be installed on servers and micro-controllers alike, so that these SDKs become a de facto standard for connecting physical and digital infrastructures to application programming interfaces (APIs) that facilitate data sharing, access to physical function, and sharing of data-driven insights.

These SDKs would also allow for the specification of executable access-control policies that can refer to existing APIs (identity management systems, distributed databases, blockchains, standardized communication protocols, and so forth) through an expressive and \emph{extensible} attribute-based language. The idea of extending domain-specific languages (DSL) in this manner has gained momentum in recent years, for example with the development and successful uptake of programming languages such as Scala~\cite{DBLP:journals/cacm/OderskyR14} and Haskell~\cite{DBLP:conf/cefp/Gibbons13}. These considerations suggest our next principle:

\begin{principle}[High Interoperability at Low Cost]
  Sharing ecosystems thrive if the software for its supporting infrastructure is lightweight, expressive, and extensible.
\end{principle}

Before we study challenges that a realization of our vision and its principles face, we first describe some initial technical work we did in this larger space and how it informed the vision set out in this paper.

\section{Our Work Done so far}
\label{section:workdonesofar}
We now describe our past and ongoing efforts in shaping this vision with R\&D partners and in developing the \frost{} technology. The latter serves as a foundation for realizing at least the first five principles above and it can play a key role in enabling the remaining principles we set out above.

\subsection{Porsche Pilot}

Our in-house experts in Cybersecurity and Blockchain were selected as winners of a competition organized by the German luxury automobile manufacturer Porsche. The prize resulted in a pilot project to bring blockchain functionality into the modern car, with a view of exploring what possibilities this would bring -- e.g.~richer user experiences, novel services or better cybersecurity of vehicles; see e.g.~\cite{LundbaekRSOS18} for a technical account of that project. 

Our team did bring blockchain functionality into a Porsche Panamera
and demonstrated not only how this approach can indeed strengthen the cybersecurity of accessing vehicles, but also how this could enable a suite of new service offerings for Porsche customers, with the resiliency that blockchain state can offer. For example, a car owner could delegate~--~through an app on a smartphone~--~to a designated delivery service access to the trunk in specific contexts such as time intervals. At
\url{https://www.youtube.com/watch?v=KvyF78RTj18} an illustrative video shows those pilot outcomes.

These capabilities were brought about by integrating a Raspberry Pi into the vehicle's internal network and the delegation and access functionality was programmed in a somewhat ad-hoc manner, given the rapid pace of the pilot project itself. Of course, we realized at that time already that we needed a more programmable means of articulating and enforcing such controls in the medium term.

\subsection{\frost{} Language}
It was this recognition and the lessons learned during that pilot project that made us research policy-based access-control in the literature. One of the authors of this paper, together with Glenn Bruns (who was at Bell Labs at the time), developed the PBel policy-composition language based on Belnap bilattices~\cite{DBLP:conf/csfw/BrunsH08,DBLP:journals/tissec/BrunsH11}. Their joint work, which began in 2006, also showed how satisfiability checkers can be used to answer a range of policy analyses that can support design, change management, and composition of policies. 

This work was therefore a natural starting point for assessing suitable approaches. There are of course many other techniques and excellent contributions to policy-based access-control, composition of policies, and policy verification in the extent literature. But we soon recognized that PBel and its analysis framework would offer many benefits if adapted judiciously to the needs of the access-control infrastructure of our envisioned sharing ecosystem. Our team then worked out these revisions, which led to the development of the \frost{} language, its compiler, and its bespoke code analysis. ``FROST'' stands for {\bf F}lexible, {\bf R}esilient, {\bf O}pen, {\bf S}ervice-Enabling, and {\bf T}rusted.

The \frost{} language expresses policies that also take on meaning over the 4-valued Belnap lattice: {\bf grant}, {\bf deny},  {\bf conflict}, and {\bf undef}. The two additional decision values {\bf conflict} and {\bf undef} were needed to adequately reflect the open nature in which policies may be composed in such ecosystems: the former captures that there is evidence for granting \emph{and} for denying a request; the latter expresses that the policy has neither evidence for granting nor for denying a request. We appreciate that the access-control community has argued for the use of even more decision values, e.g.~in the standard XACML~3.0 of the Organization for the Advancement of Structured Information Standards.
But our engagement with business partners
made us move additional values and their complexity into constraints whose truth values influence decisions by triggering rules within policies. 

Like PBel, the \frost{} language is rule-based and its composition
primitive is a ${\bf case}$-statement in which guards capture decision
scenarios of one or more sub-policies. \frost{} is also
\emph{attribute-based} in that its atomic conditions are formed from
relations of attributes. The latter accommodates integration with
other technology through APIs; e.g.~we have developed such an API for
the Security Assertion Markup Language. Attributes and their
verification are a powerful tool. For example, an attribute {\bf
  subject} can resolve to a more complex meaning such as multiple
agents who issue a request based on a majority vote.\footnote{We
  appreciate that access-control \emph{models} such as UCON
  \cite{UCON} distinguish between subjects and subject
  attributes. FROST allows the use of more abstract DSLs that compile
  into FROST and that do make such distinctions.} 

Let us illustrate this with examples of policy and policy composition
in \frost{}. The policy in Figure~\ref{fig:policy} consists of a sole
rule that {\bf grant}s the access when its condition is true and
returns {\bf undef} otherwise. That condition is a conjunction of constraints that provide mappings to the object, subject, and action of the access request and where additional conditions such as time windows constrain that access. We note that the semantics of these mappings is implementation-specific. For example, evaluating whether the subject is the daughter of the owner may be done through multi-factor authentication (live AI for face recognition, biometrics of the driver seat, and so forth).
\begin{figure}
    \centering
    \begin{eqnarray*}
    {\bf grant\ if} &{}& ({\bf object}\ == vehicle)\,\&\&\\
    {}&{}&  ({\bf subject}\ == vehicle.owner.daughter)\,\&\&\\
    {}&{}& ({\bf action}\ == driveVehicle)\,\&\&\\
    {}&{}& (owner.daughter.driversLicense == valid)\,\&\&\\
    {}&{}& (0900 \leq localTime)\,\&\&\\
    {}&{}& (localTime\leq 2000)
    \end{eqnarray*}
    \caption{Example \frost{} policy that allows the daughter of a vehicle owner to drive that vehicle under specific conditions pertaining to time windows, valid driving license, and potential authentication implicit in the ``{\bf subject}'' condition}
    \label{fig:policy}
\end{figure}

Figure~\ref{fig:policy-composition} shows a ${\bf case}$-statement, a
priority composition of policies $P$ and $Q$. \frost{} cases are of
the form $[ guard\colon policy]$ where $guard$ is either ${\bf true}$
or a conjunction of expressions of form
``$policy\ {\bf eval}\ decision$''
saying that $policy$ evaluates to a $decision$, an element of the set
$\{{\bf grant}, {\bf deny}, {\bf conflict}, {\bf undef}\}$.
The composition in Figure~\ref{fig:policy-composition} therefore denies if $P$ reports a {\bf conflict}, returns the decision of $Q$ if $P$ reports {\bf undef}, and returns the decision of $P$ in all other cases. Deep embeddings of \frost{} into other languages such as Haskell would also support naming of conditions and policies, as well as function declarations and invocations whose return types are \frost{} conditions or policies.
\begin{figure}
\centering
\begin{eqnarray*}
{} &{}& {\bf case}\ \{ \\
{} &{}& \ \ \ \ [P\ {\bf eval}\ {\bf conflict}\ \colon {\bf deny} ] \\
{} &{}& \ \ \ \ [P\ {\bf eval}\ {\bf undef}\ \colon Q ] \\
{} &{}& \ \ \ \ [{\bf true}\colon P] \\
{} &{}& \}
\end{eqnarray*}
    \caption{Example policy composition specified with a ${\bf
        case}$-statement: this encodes a binary operator $P >\!> Q$
      where policy parameter $P$ has priority over policy parameter $Q$}
    \label{fig:policy-composition}
\end{figure}

The \frost{} language also accommodates obligations, in an approach similar to one of~\cite{DBLP:conf/sacmat/LiWQBRLL09}; but other obligation semantics are consistent with the \frost{} language as well. We have not yet implemented obligations, but are interested in exploring their enforcement aspects (e.g.~for payments) through an open \frost{} network that makes use of DLT.

Finally, a \frost{} policy is compiled into two Boolean circuits that capture the 4-valued decision space and whose atomic expressions are actually atomic conditions  occurring in the compiled \frost{} policies. It is these circuits that will be executed in the access-control architecture, making this execution environment simple and efficient enough for micro-controller environments as well.
We refer to the \frost{} Yellow Paper~\cite{XainFROSTYellowPaper} for further details.

\subsection{Access-control Architecture}

The access-control architecture for \frost{} is akin to that of XACML, making use of policy decision points, policy enforcement points, policy administration points, policy information points, and so forth~\cite{XainFROSTYellowPaper}.

We developed a reference client for this architecture, at present with minimal functionality, that can also run in a micro-controller environment. One thing we realized in its development is that the policy decision point can be very generic but the code for the policy enforcement point may vary as a function of the host environment (e.g.~whether it is a micro-controller or a server or whether the decisions would be enforced in a moving vehicle).

\subsection{Delegation Chains}

Our vision foresees that actors can program their own sharing ecosystem. One aspect of that is the articulation and enforcement of a \emph{delegation graph} that expresses the intent of actors in regard to the sharing of their assets.

So far, we designed cybersecurity protocols that an asset owner can
initiate in order to build up a delegation \emph{chain} of actors,
where the chain starts with the asset owner and ends with the software
agent that controls access to that asset. Intermediate actors are then
permitted to submit their own policies. The overall policy executed on
the asset is the composition of these policies, but where the choice
of the composition operator is controlled by the asset owner. In
\frost{}, we decouple syntax for conditions and policies from syntax
and semantics of delegation; this simplifies reasoning and allows for
the adoption of other approaches to delegation.

\subsection{Infineon Project}

We also entered a partnership with Infineon Technologies, after we demonstrated successfully in a minimal viable product how the capabilities of the Porsche pilot could be realized not on a Raspberry Pi but on an Electronic Control Unit that is part of the vehicle's product platform, an Aurix\texttrademark{} micro-controller:

\medskip\par\noindent
\url{www.infineon.com/cms/en/about-infineon/press/press-releases/2018/INFATV201810-005.html}
\medskip\par

The execution of that project made us appreciate the extent and types of demands that processing and storage constraints of micro-processors place on realizing use cases of our vision of sharing ecosystems.

Through this partnership, we hope to be able to mature the \frost{} technology so that it will become production-ready for integration in the automotive sector, where lead times for such technology changes range from 3-5 years. A key aspect will be to integrate our technical approach with a de facto software standard, Autosar,  for interfacing with the internal network of vehicles.

\section{Challenges for Our Vision}
\label{section:challenges}

In this section, we outline some of the key challenges that have to be addressed in order to realize our vision.

\subsection{From Closed to Open Ecosystems}

From our experience, it seems prudent to build up sharing systems and their supporting technologies bottom-up, rather than top-down. The latter approach would first build an entire economic platform and has then the problem of drawing parties into using that platform as a marketplace for access to data, physical resources, and data-driven insights.

A bottom-up approach ensures that potential users of such a platform are actively involved in defining and shaping the technical aspects of that platform. This also seems to be a  better fit for commercial parties, since they are interested in trying this first out in a (semi-)closed environment in order to fully control with whom to engage in sharing interactions. 

An important part of our vision is an open sharing network, populated by many sharing biotopes that may interact with each other.
One challenge faced here is to build up a sufficient number of biotopes to allow for network effects, and that at present the setup of a biotope requires social and technical on-boarding that should ideally be supported with as much automation as possible.

\subsection{Building a Technical Community}

Almost all of our code development for \frost{} will be open-sourced and  its license models are meant to encourage open-source contributions so that others can freely use such code, also for further derived Intellectual Property that they may wish to exploit commercially. 

One challenge in this is to build up a community of code developers that can make such contributions. The Rust programming language has its own ecosystem of developers but this is still a fairly small community, which limits the amount of qualified programmers as contributors and as potential recruitment targets. In automotive, use of C instead of Rust is currently preferred as this makes certification and compliance activities faster and cheaper.

We hope our Haskell-based DSL will help draw some interest from developers (and potentially academics) at the intersection of the access-control and functional programming communities.

\subsection{Geographical Dynamics}

Our vision and its supporting technology are about infrastructure and, as such, are meant to be global in reach. 
One challenge, certainly within Europe, is that moving from a pilot project to a scalable production case may take 3-5 years~--~and in some sectors even longer than that (e.g.~automotive platforms last 10-15 years). This is a challenge for start-ups such as XAIN AG. We therefore are keen to also engage with partners who can integrate our technology in large production systems in 2019 or 2020 already.

\subsection{Optimal Trade-offs}

Access-control was invented as a means for making systems more secure. In sharing ecosystems, security and privacy demands are certainly very important. But trade-offs may have to be made. For example, it is not feasible to make an electronic control unit within a car into a full node of some blockchain, as the storage and compute resources of these devices won't allow for this. 

This means that we need to develop hybrid approaches, for example one in which such an Electronic Control Unit only stores partial state of a blockchain that is pertinent to its own security state. Working this out in practice, with an appropriate threat and mitigation model, poses a challenge.

\section{Engaging Academic Community}
\label{section:academic}

The academic community of access-control has indubitably a tremendous wealth of knowledge, tools, and general expertise to offer in order to support the shaping and realization of our vision. We here sketch some of the ways in which this community may help. But we want to stress that members of that community are best placed to come forward with ideas and proposals for how they can engage in this long-term project. We therefore are eager to hear from that community directly about what it thinks it could offer here.

Below, we feature some topics whose development would be of value to realizing our vision and that we think would be of interest to that academic community:

\subsection{Policy Life Cycle Management}

A trusted life cycle of an access-control policy and its potential updates and retirement are important. How this life cycle is specified and enforced can be seen as an aspect of policy administration. We seek models for this life cycle (e.g.~automata-based models of permitted administration actions) that can be stored and executed on resource-constrained micro-controllers as well as on enterprise servers.

\subsection{Policy Integrity}

How best can we ascertain the integrity of an access-control policy, not just its provenance but also its freshness? DLT solutions suggest themselves, so that integrity tags of policies and their update history may be recorded on a distributed ledger. We have some initial concepts in that space, using a block\emph{tree} rather than a blockchain, to reflect that many of these policies are independent of each other and so state can be sharded for higher scalability. This approach is based on a similar proposal for Bitcoin found at 

\medskip\par\noindent
\url{https://lists.linuxfoundation.org/pipermail/bitcoin-dev/2014-March/004797.html}
\par\noindent

\subsection{Policy Verification}

We realize that the academic community has done a lot of work on the formal verification of access-control policies and their administration. Here, we seek tools that make use of such established methods and techniques, e.g.~Satisfiability Modulo Theories solvers (see e.g.~\cite{DBLP:conf/tacas/MouraB08}). We already have a good understanding of the first analyses we want to support (e.g.~dead code analysis on policies). But we certainly stand to benefit from making use of existing tools of the community. And we think that \frost{} can offer interesting formal verification problems and impact opportunities to the academic community.

\subsection{Policy Privacy}

Privacy considerations are of paramount importance in many use cases. We would welcome collaborations with the academic community to ensure that \frost{} can be adapted to varying privacy needs. For example, our current delegation protocol only hides policies of delegatees from other delegatees and not from the device that stores and executes the composed policy. And this limited privacy guarantee is based on encryption, which may be too weak for some use cases that require higher privacy levels~--~e.g.~as those associated with the ``Right to be Forgotten'' in the EU's General Data Protection Regulation (GDPR).

It is clear that the academic research community has a lot to offer in understanding which architectures for access-control and its administration can support very strict privacy demands such as those imposed by the GDPR. As the report by the French CNIL, found at 

\medskip\par
\url{https://www.cnil.fr/sites/default/files/atoms/files/blockchain.pdf}
\medskip\par

\noindent suggests, it is challenging to make blockchains GDPR-com\-pliant but there seem to be ways forward in achieving this. We are particularly interested in solutions where blockchains (or blocktrees) manage the administrative and security state of access-control policies and their composition, rather than managing the history of financial transactions.

\subsection{Access-Control on a Chip}

In the medium to long term, we are interested in putting such complex access-control architectures on a chip that may host a secure database (which e.g.\ may store partial blocktree state). We are therefore eager to engage with academics who offer expertise in chip design for access-control.

\subsection{App Store}

Part of our vision is that others can use this sharing infrastructure in order to program applications on top of it, and where these applications make use of the sharing mechanisms of that infrastructure. For example, one could imagine an Artificial Intelligence app that is installed in vehicles that perform local reinforcement learning but where the models are updated across these vehicles for better accuracy and learning speed; and where \frost{} would facilitate the access restrictions but also the incentives for such sharing. We are thus also interested in what the academic community can offer in that regard. Topics of interest here seem to be algorithmic foundations for writing such apps, interface technology that can connect such apps with our infrastructure, user interfaces for specifying the access controls pertinent to such an app, to name a few.

\subsection{Academic Outreach}

We are keen to engage with the academic community through established means. For example, we would be prepared to organize a technical workshop dedicated to our vision~--~perhaps affiliated with an established conference. We would also welcome academics to visit our headquarters in Berlin, Germany, for a face-to-face chat about our vision and ongoing work and possible research collaborations.

\section{Related Work}
\label{section:relatedwork}
In this section, we feature related work that pertains to the vision we have set out in this paper. We group this work into four themes rather than giving a many-to-many mapping of such work to the nine principles of our vision.

\subsection{Sharing and Collaboration}
First, we relate our vision to research on access-control for sharing and collaboration.
In \cite{DBLP:conf/icsoc/RissanenBS09}, XACML is used to refine existing enterprise access-control systems so that they allow for collaboration in distributed environments, also between different organizations.
Role-based access-control is combined with attribute-based access-control in \cite{DBLP:journals/wpc/LoYG15} to improve the security of multi-tenant cloud environments.
Multi-tenancy architectures for Software as a Service can benefit from support for sub-tenancies. But this requires secure management of such complexity. An access-control model based on Administrative RBAC is proposed in \cite{DBLP:journals/fcsc/ZuoXQZ17} in order to address this.
In \cite{DBLP:conf/cloudcomp/LiZJX16}, key distribution is combined with the authorization of users and a procedure for data sharing to allow for secure data sharing without the need of a trusted third party.
Open challenges for access-control of jointly owned or shared documents are discussed in~\cite{DBLP:conf/sacmat/SquicciariniRZ18}, a solution for the cloud environment is presented in~\cite{DBLP:conf/sacmat/SorienteKRMC15}, and a data-centric cloud solution is offered in~\cite{DBLP:conf/sacmat/PasquierBSE16}. Cross-organizational tracking of assets also is in need of access-control in order to avoid unintended confidentiality leaks~\cite{DBLP:conf/sacmat/HanZGB15}.  Access-control for the provisioning of attributes and policies in collaborative network environments is developed in~\cite{DBLP:conf/sacmat/Rubio-MedranoZD15}.
Work that uses domain knowledge and semantic relations between data and data usage, see e.g.~\cite{DBLP:conf/sacmat/PaciZ15}, can improve policies for sharing and their enforcement. 
The sound and resilient recovery of policy state from missing attribute information, see e.g.~\cite{XainFROSTYellowPaper,DBLP:conf/sacmat/CramptonMZ15}, is important in sharing ecosystems.
Relationship-based access-control~\cite{DBLP:conf/codaspy/BrunsFSH12,DBLP:conf/sacmat/CramerPZ15,DBLP:conf/sacmat/MehreganF16,DBLP:journals/jcs/RoscheisenW97}, its policy negotiation~\cite{DBLP:conf/sacmat/MehreganF16}, and its modelling languages~\cite{DBLP:conf/sacmat/PasarellaL17} seem to fit well user experiences in sharing ecosystems.

\subsection{Safety, Resiliency, and Trust}
Second, we discuss work regarding the aspects of safety, resiliency, and trust.
Access-control for sharing displays in a vehicle can ensure the safety of display sharing~\cite{DBLP:conf/sacmat/GanselSGFDRM14}.
In \cite{DBLP:journals/access/WangZZ18a}, decentralized storage and blockchain technology are combined with attribute-based encryption to provide data privacy and fine-grained access-control while avoiding a single point of failure.
Attestable trust anchors on clients can help with usage control of data across devices~\cite{DBLP:conf/sacmat/WagnerBB18}. Combining reputation anchors with obligation handling~\cite{DBLP:conf/sacmat/BotelhoPN13} or with DLT mechanisms such as obligation chains~\cite{DBLP:conf/sacmat/PietroSSW18} can make trust more resilient in open systems.
Hardware isolation of the access-control functionality from the normal application layers, e.g.~as done in~\cite{DBLP:conf/sacmat/HaiderOLDD17}, is of great interest for secure sharing technology.
 Access-control enforcement itself carries risks that may be mitigated through run-time tracking of risk estimates~\cite{DBLP:conf/sacmat/PetraccaCSJ17}. 
The need for access-control protection measures at different abstraction layers of systems is well recognized (see e.g.~\cite{DBLP:conf/sacmat/Sadeghi13}). And the increasingly distributed and intelligent nature of connected systems brings both research challenges and opportunities for policy-based system management~\cite{DBLP:conf/sacmat/CaloVB17}.

\subsection{Compute Constraints}
Third, we discuss work pertaining to the compute constraints in IoT systems.
Access-control is an important aspect of vehicular cloud environments~\cite{DBLP:conf/sacmat/GuptaS18}.
The challenge of compute demands for secure access-control within vehicle systems is recognized in \cite{DBLP:journals/network/XueHMWHY18}, a fog-to-cloud architecture is proposed to address this limitation while also reducing latency of cloud-based solutions.
The Policy Machine \cite{DBLP:journals/jsa/FerraioloAG11} is a framework that can host access-control enforcement with minimal code requirements, a generic mechanism, and expressiveness for refining constraints in high-assurance domains.

\subsection{Policy Analysis and Synthesis}
Fourth, we briefly relate to work on synthesis or analysis of access-control policies. %
The high degree of dynamics in IoT networks may benefit from use of event-driven policies~\cite{DBLP:conf/sacmat/PietroSSW18} and a less static generation and enforcement of access-control policies (e.g.~through machine learning)~\cite{DBLP:conf/sacmat/CaloVCBLC18}. 
Machine learning can help with converting natural language policies into formal access-control policies, e.g.~through the identification of attributes in~\cite{DBLP:conf/sacmat/AlohalyTB18}.
Policy analysis tools can  identify excessive privileges and redundancies~\cite{DBLP:conf/sacmat/ChariMPT13}, establish privacy properties~\cite{DBLP:conf/sacmat/ChowdhuryGNRBDJW13}, and improve revocation schemes for delegation chains~\cite{DBLP:conf/sacmat/CramerAH15}.

\section{Conclusions}
\label{section:conclusion}

This paper articulated our ambitious vision of technology for sharing ecosystems and its relevance to the access-control research community. Furthermore, we discussed exploratory and more concrete work that we already  undertook towards realizing this vision. Then we talked about  several challenges that we face in bringing this vision to fruition. We pointed out that we can benefit greatly from the broad and deep expertise of the academic community in access-control in overcoming these challenges in the medium to long term. We also reviewed research from the academic literature in order to illustrate pertinent connections between that extant work and our socio-technical vision. 

As we said already, we would be very grateful for gaining any feedback on this vision, receiving any expressions of interest in research collaborations, and in obtaining any contributions to our open-source project for the \frost{} access-control and infrastructure technology.

\section{Acknowledgments}
The second author would like to acknowledge 
 UK EPSRC grant EP/N020030/1, the UK EPSRC grant EP/N023242/1, as well as the LRF supported project DDIP-IoT.


{\small

\begin{thebibliography}{10}

\bibitem{DBLP:conf/sacmat/AlohalyTB18}
M.~Alohaly, H.~Takabi, and E.~Blanco.
\newblock {A} {D}eep {L}earning {A}pproach for {E}xtracting {A}ttributes of
  {ABAC} {P}olicies.
\newblock In {\em Proceedings of the 23nd {ACM} on Symposium on Access Control
  Models and Technologies, {SACMAT} 2018, Indianapolis, IN, USA, June 13-15,
  2018}, pages 137--148, 2018.

\bibitem{DBLP:conf/sacmat/BotelhoPN13}
B.~A.~P. Botelho, D.~G. Pelluzi, and E.~T. Nakamura.
\newblock {A} versatile access control implementation: secure box.
\newblock In {\em 18th {ACM} Symposium on Access Control Models and
  Technologies, {SACMAT} '13, Amsterdam, The Netherlands, June 12-14, 2013},
  pages 249--252, 2013.

\bibitem{DBLP:conf/codaspy/BrunsFSH12}
G.~Bruns, P.~W.~L. Fong, I.~Siahaan, and M.~Huth.
\newblock {R}elationship-based access control: its expression and enforcement
  through hybrid logic.
\newblock In {\em Second {ACM} Conference on Data and Application Security and
  Privacy, {CODASPY} 2012, San Antonio, TX, USA, February 7-9, 2012}, pages
  117--124, 2012.

\bibitem{DBLP:conf/csfw/BrunsH08}
G.~Bruns and M.~Huth.
\newblock {A}ccess-{C}ontrol {P}olicies via {B}elnap {L}ogic: {E}ffective and
  {E}fficient {C}omposition and {A}nalysis.
\newblock In {\em {P}roceedings of the 21st {IEEE} {C}omputer {S}ecurity
  {F}oundations {S}ymposium, {CSF} 2008, {P}ittsburgh, {P}ennsylvania, {USA},
  23-25 {J}une 2008}, pages 163--176, 2008.

\bibitem{DBLP:journals/tissec/BrunsH11}
G.~Bruns and M.~Huth.
\newblock {A}ccess control via {B}elnap logic: {I}ntuitive, expressive, and
  analyzable policy composition.
\newblock {\em {ACM} Trans. Inf. Syst. Secur.}, 14(1):9:1--9:27, 2011.

\bibitem{DBLP:conf/sacmat/CaloVB17}
S.~B. Calo, D.~C. Verma, and E.~Bertino.
\newblock {D}istributed {I}ntelligence: {T}rends in the {M}anagement of
  {C}omplex {S}ystems.
\newblock In {\em Proceedings of the 22nd {ACM} on Symposium on Access Control
  Models and Technologies, {SACMAT} 2017, Indianapolis, IN, USA, June 21-23,
  2017}, pages 1--7, 2017.

\bibitem{DBLP:conf/sacmat/CaloVCBLC18}
S.~B. Calo, D.~C. Verma, S.~Chakraborty, E.~Bertino, E.~Lupu, and G.~H.
  Cirincione.
\newblock {S}elf-{G}eneration of {A}ccess {C}ontrol {P}olicies.
\newblock In {\em Proceedings of the 23nd {ACM} on Symposium on Access Control
  Models and Technologies, {SACMAT} 2018, Indianapolis, IN, USA, June 13-15,
  2018}, pages 39--47, 2018.

\bibitem{DBLP:conf/osdi/CastroL99}
M.~Castro and B.~Liskov.
\newblock {P}ractical {B}yzantine {F}ault {T}olerance.
\newblock In {\em {P}roceedings of the {T}hird {USENIX} {S}ymposium on
  {O}perating {S}ystems {D}esign and {I}mplementation ({OSDI}), {N}ew
  {O}rleans, {L}ouisiana, {USA}, {F}ebruary 22-25, 1999}, pages 173--186, 1999.

\bibitem{DBLP:conf/sacmat/ChariMPT13}
S.~Chari, I.~Molloy, Y.~Park, and W.~Teiken.
\newblock {E}nsuring continuous compliance through reconciling policy with
  usage.
\newblock In {\em 18th {ACM} Symposium on Access Control Models and
  Technologies, {SACMAT} '13, Amsterdam, The Netherlands, June 12-14, 2013},
  pages 49--60, 2013.

\bibitem{DBLP:journals/corr/Micali16}
J.~Chen and S.~Micali.
\newblock {ALGORAND}.
\newblock {\em CoRR}, abs/1607.01341, 2016.

\bibitem{XainFROSTYellowPaper}
K.~Cheung, M.~R.~A. Huth, L.~M. Kirk, L.~Lundb{\ae}k, R.~Marques, and
  J.~Petsche.
\newblock The {FROST} {L}anguage: {A} trusted and user-centric access control
  language: {E}nabling delegation of fine-grained policies in shared
  ecosystems.
\newblock Yellow Paper freely available online at
  \url{https://xain.foundation}, October 2018.
\newblock Version 0.9.

\bibitem{DBLP:conf/sacmat/ChowdhuryGNRBDJW13}
O.~Chowdhury, A.~Gampe, J.~Niu, J.~von Ronne, J.~Bennatt, A.~Datta, L.~Jia, and
  W.~H. Winsborough.
\newblock {P}rivacy promises that can be kept: a policy analysis method with
  application to the {HIPAA} privacy rule.
\newblock In {\em 18th {ACM} Symposium on Access Control Models and
  Technologies, {SACMAT} '13, Amsterdam, The Netherlands, June 12-14, 2013},
  pages 3--14, 2013.

\bibitem{DBLP:conf/sacmat/CramerAH15}
M.~Cramer, D.~A. Ambrossio, and P.~V. Hertum.
\newblock {A} {L}ogic of {T}rust for {R}easoning about {D}elegation and
  {R}evocation.
\newblock In {\em Proceedings of the 20th {ACM} Symposium on Access Control
  Models and Technologies, Vienna, Austria, June 1-3, 2015}, pages 173--184,
  2015.

\bibitem{DBLP:conf/sacmat/CramerPZ15}
M.~Cramer, J.~Pang, and Y.~Zhang.
\newblock {A} {L}ogical {A}pproach to {R}estricting {A}ccess in {O}nline
  {S}ocial {N}etworks.
\newblock In {\em Proceedings of the 20th {ACM} Symposium on Access Control
  Models and Technologies, Vienna, Austria, June 1-3, 2015}, pages 75--86,
  2015.

\bibitem{DBLP:conf/sacmat/CramptonMZ15}
J.~Crampton, C.~Morisset, and N.~Zannone.
\newblock {O}n {M}issing {A}ttributes in {A}ccess {C}ontrol:
  {N}on-deterministic and {P}robabilistic {A}ttribute {R}etrieval.
\newblock In {\em Proceedings of the 20th {ACM} Symposium on Access Control
  Models and Technologies, Vienna, Austria, June 1-3, 2015}, pages 99--109,
  2015.

\bibitem{DBLP:conf/tacas/MouraB08}
L.~M. de~Moura and N.~Bj{\o}rner.
\newblock {Z3}: {A}n {E}fficient {SMT} {S}olver.
\newblock In {\em {T}ools and {A}lgorithms for the {C}onstruction and
  {A}nalysis of {S}ystems, 14th {I}nternational {C}onference, {TACAS} 2008,
  {H}eld as {P}art of the {J}oint {E}uropean {C}onferences on {T}heory and
  {P}ractice of {S}oftware, {ETAPS} 2008, {B}udapest, {H}ungary, {M}arch
  29-{A}pril 6, 2008. {P}roceedings}, pages 337--340, 2008.

\bibitem{DBLP:conf/crypto/DworkN92}
C.~Dwork and M.~Naor.
\newblock {P}ricing via {P}rocessing or {C}ombatting {J}unk {M}ail.
\newblock In {\em {A}dvances in {C}ryptology - {CRYPTO} '92, 12th {A}nnual
  {I}nternational {C}ryptology {C}onference, {S}anta {B}arbara, {C}alifornia,
  {USA}, {A}ugust 16-20, 1992, {P}roceedings}, pages 139--147, 1992.

\bibitem{DBLP:journals/jsa/FerraioloAG11}
D.~F. Ferraiolo, V.~Atluri, and S.~I. Gavrila.
\newblock The policy machine: {A} novel architecture and framework for access
  control policy specification and enforcement.
\newblock {\em Journal of Systems Architecture - Embedded Systems Design},
  57(4):412--424, 2011.

\bibitem{DBLP:conf/sacmat/GanselSGFDRM14}
S.~Gansel, S.~Schnitzer, A.~Gilbeau{-}Hammoud, V.~Friesen, F.~D{\"{u}}rr,
  K.~Rothermel, and C.~Maih{\"{o}}fer.
\newblock {A}n access control concept for novel automotive {HMI} systems.
\newblock In {\em 19th {ACM} Symposium on Access Control Models and
  Technologies, {SACMAT} '14, London, ON, Canada - June 25 - 27, 2014}, pages
  17--28, 2014.

\bibitem{DBLP:conf/cefp/Gibbons13}
J.~Gibbons.
\newblock {F}unctional {P}rogramming for {D}omain-{S}pecific {L}anguages.
\newblock In {\em {C}entral {E}uropean {F}unctional {P}rogramming {S}chool -
  5th {S}ummer {S}chool, {CEFP} 2013, {C}luj-{N}apoca, {R}omania, July 8-20,
  2013, {R}evised {S}elected {P}apers}, pages 1--28, 2013.

\bibitem{DBLP:conf/sacmat/GuptaS18}
M.~Gupta and R.~S. Sandhu.
\newblock {A}uthorization {F}ramework for {S}ecure {C}loud {A}ssisted
  {C}onnected {C}ars and {V}ehicular {I}nternet of {T}hings.
\newblock In {\em Proceedings of the 23nd {ACM} on Symposium on Access Control
  Models and Technologies, {SACMAT} 2018, Indianapolis, IN, USA, June 13-15,
  2018}, pages 193--204, 2018.

\bibitem{DBLP:conf/sacmat/HaiderOLDD17}
S.~K. Haider, H.~Omar, I.~A. Lebedev, S.~Devadas, and M.~van Dijk.
\newblock {L}everaging {H}ardware {I}solation for {P}rocess {L}evel {A}ccess
  {C}ontrol {\&} {A}uthentication.
\newblock In {\em Proceedings of the 22nd {ACM} on Symposium on Access Control
  Models and Technologies, {SACMAT} 2017, Indianapolis, IN, USA, June 21-23,
  2017}, pages 133--141, 2017.

\bibitem{DBLP:conf/sacmat/HanZGB15}
W.~Han, Y.~Zhang, Z.~Guo, and E.~Bertino.
\newblock {F}ine-{G}rained {B}usiness {D}ata {C}onfidentiality {C}ontrol in
  {C}ross-{O}rganizational {T}racking.
\newblock In {\em Proceedings of the 20th {ACM} Symposium on Access Control
  Models and Technologies, Vienna, Austria, June 1-3, 2015}, pages 135--145,
  2015.

\bibitem{DBLP:journals/pacmpl/0002JKD18}
R.~Jung, J.~Jourdan, R.~Krebbers, and D.~Dreyer.
\newblock {R}ust{B}elt: securing the foundations of the {R}ust programming
  language.
\newblock {\em {PACMPL}}, 2 ({POPL}):66:1--66:34, 2018.

\bibitem{DBLP:conf/sacmat/LiWQBRLL09}
N.~Li, Q.~Wang, W.~H. Qardaji, E.~Bertino, P.~Rao, J.~Lobo, and D.~Lin.
\newblock {A}ccess control policy combining: theory meets practice.
\newblock In {\em 14th {ACM} {S}ymposium on {A}ccess {C}ontrol {M}odels and
  {T}echnologies, {SACMAT} 2009, {S}tresa, {I}taly, {J}une 3-5, 2009,
  {P}roceedings}, pages 135--144, 2009.

\bibitem{DBLP:conf/cloudcomp/LiZJX16}
Z.~Li, M.~Zhao, H.~Jiang, and Q.~Xu.
\newblock Data sharing with fine-grained access control for multi-tenancy cloud
  storage system.
\newblock In {\em Cloud Computing, Security, Privacy in New Computing
  Environments - 7th International Conference, CloudComp 2016, and First
  International Conference, {SPNCE} 2016, Guangzhou, China, November 25-26, and
  December 15-16, 2016, Proceedings}, pages 123--132, 2016.

\bibitem{DBLP:journals/wpc/LoYG15}
N.~Lo, T.~C. Yang, and M.~Guo.
\newblock An attribute-role based access control mechanism for multi-tenancy
  cloud environment.
\newblock {\em Wireless Personal Communications}, 84(3):2119--2134, 2015.

\bibitem{xainpokw}
L.-N. Lundb{\ae}k, D.~J. Beutel, M.~Huth, S.~Jackson, and L.~Kirk.
\newblock {P}roof of {K}ernel {W}ork: {A} {R}esilient \& {S}calable
  {B}lockchain {C}onsensus {A}lgorithm for {D}ynamic {L}ow-{E}nergy {N}etworks.
\newblock {\em Yellow Paper}, v1.3, 2018.

\bibitem{LundbaekRSOS18}
L.-N. Lundb{\ae}k, D.~J. Beutel, M.~Huth, S.~Jackson, L.~Kirk, and R.~Steiner.
\newblock {P}roof of {K}ernel {W}ork: a democratic low-energy consensus for
  distributed access-control protocols.
\newblock {\em Royal Society Open Science}, 5(8), 2018.

\bibitem{IoT15}
J.~Manyika, M.~Chui, P.~Bisson, J.~Woetzel, R.~Dobbs, J.~Bughin, and D.~Aharon.
\newblock {U}nlocking the potential of the {I}nternet of {T}hings, {M}c{K}insey
  {R}eport, published online, June 2015.

\bibitem{DBLP:conf/sacmat/MehreganF16}
P.~Mehregan and P.~W.~L. Fong.
\newblock {P}olicy {N}egotiation for {C}o-owned {R}esources in
  {R}elationship-{B}ased {A}ccess {C}ontrol.
\newblock In {\em Proceedings of the 21st {ACM} on Symposium on Access Control
  Models and Technologies, {SACMAT} 2016, Shanghai, China, June 5-8, 2016},
  pages 125--136, 2016.

\bibitem{AutoMcKinsey16}
D.~Mohr, H.-W. Kaas, P.~Gao, D.~Wee, and T.~M{\"o}ller.
\newblock {A}utomotive revolution~--~perspective towards 2030, published
  online, January 2016.

\bibitem{nakamoto08}
S.~Nakamoto.
\newblock {B}itcoin: {A} {P}eer-to-{P}eer {E}lectronic {C}ash {S}ystem, May
  2008.
\newblock Published under pseudonym.

\bibitem{Nicolescu2018}
R.~Nicolescu, M.~Huth, P.~Radanliev, and D.~D. Roure.
\newblock {M}apping the values of {I}o{T}.
\newblock {\em Journal of Information Technology}, Mar 2018.

\bibitem{DBLP:journals/cacm/OderskyR14}
M.~Odersky and T.~Rompf.
\newblock {U}nifying functional and object-oriented programming with {S}cala.
\newblock {\em Commun. {ACM}}, 57(4):76--86, 2014.

\bibitem{DBLP:conf/sacmat/PaciZ15}
F.~Paci and N.~Zannone.
\newblock {P}reventing {I}nformation {I}nference in {A}ccess {C}ontrol.
\newblock In {\em Proceedings of the 20th {ACM} Symposium on Access Control
  Models and Technologies, Vienna, Austria, June 1-3, 2015}, pages 87--97,
  2015.

\bibitem{UCON}
J.~Park and R.~S. Sandhu.
\newblock The {UCON}\textsubscript{ABC} usage control model.
\newblock {\em {ACM} Trans. Inf. Syst. Secur.}, 7(1):128--174, 2004.

\bibitem{DBLP:conf/sacmat/PasarellaL17}
E.~Pasarella and J.~Lobo.
\newblock {A} {D}atalog {F}ramework for {M}odeling {R}elationship-based
  {A}ccess {C}ontrol {P}olicies.
\newblock In {\em Proceedings of the 22nd {ACM} on Symposium on Access Control
  Models and Technologies, {SACMAT} 2017, Indianapolis, IN, USA, June 21-23,
  2017}, pages 91--102, 2017.

\bibitem{DBLP:conf/sacmat/PasquierBSE16}
T.~F.~J. Pasquier, J.~Bacon, J.~Singh, and D.~M. Eyers.
\newblock {D}ata-{C}entric {A}ccess {C}ontrol for {C}loud {C}omputing.
\newblock In {\em Proceedings of the 21st {ACM} on Symposium on Access Control
  Models and Technologies, {SACMAT} 2016, Shanghai, China, June 5-8, 2016},
  pages 81--88, 2016.

\bibitem{DBLP:conf/sacmat/PetraccaCSJ17}
G.~Petracca, F.~Capobianco, C.~Skalka, and T.~Jaeger.
\newblock {O}n {R}isk in {A}ccess {C}ontrol {E}nforcement.
\newblock In {\em Proceedings of the 22nd {ACM} on Symposium on Access Control
  Models and Technologies, {SACMAT} 2017, Indianapolis, IN, USA, June 21-23,
  2017}, pages 31--42, 2017.

\bibitem{DBLP:conf/sacmat/PietroSSW18}
R.~D. Pietro, X.~Salleras, M.~Signorini, and E.~Waisbard.
\newblock {A} blockchain-based {T}rust {S}ystem for the {I}nternet of {T}hings.
\newblock In {\em Proceedings of the 23nd {ACM} on Symposium on Access Control
  Models and Technologies, {SACMAT} 2018, Indianapolis, IN, USA, June 13-15,
  2018}, pages 77--83, 2018.

\bibitem{DBLP:conf/icsoc/RissanenBS09}
E.~Rissanen, D.~Brossard, and A.~Slabbert.
\newblock Distributed access control management - {A} {XACML}-based approach.
\newblock In {\em Service-Oriented Computing, 7th International Joint
  Conference, ICSOC-ServiceWave 2009, Stockholm, Sweden, November 24-27, 2009.
  Proceedings}, pages 639--640, 2009.

\bibitem{DBLP:journals/jcs/RoscheisenW97}
M.~R{\"{o}}scheisen and T.~Winograd.
\newblock {A} {N}etwork-{C}entric {D}esign for {R}elationship-{B}ased
  {S}ecurity and {A}ccess {C}ontrol.
\newblock {\em Journal of Computer Security}, 5(3):249--254, 1997.

\bibitem{DBLP:conf/sacmat/Rubio-MedranoZD15}
C.~E. Rubio{-}Medrano, Z.~Zhao, A.~Doup{\'{e}}, and G.~Ahn.
\newblock {F}ederated {A}ccess {M}anagement for {C}ollaborative {N}etwork
  {E}nvironments: {F}ramework and {C}ase {S}tudy.
\newblock In {\em Proceedings of the 20th {ACM} Symposium on Access Control
  Models and Technologies, Vienna, Austria, June 1-3, 2015}, pages 125--134,
  2015.

\bibitem{DBLP:conf/sacmat/Sadeghi13}
A.~Sadeghi.
\newblock {M}obile security and privacy: the quest for the mighty access
  control.
\newblock In {\em 18th {ACM} Symposium on Access Control Models and
  Technologies, {SACMAT} '13, Amsterdam, The Netherlands, June 12-14, 2013},
  pages 1--2, 2013.

\bibitem{DBLP:conf/sacmat/SorienteKRMC15}
C.~Soriente, G.~O. Karame, H.~Ritzdorf, S.~Marinovic, and S.~Capkun.
\newblock {C}ommune: {S}hared {O}wnership in an {A}gnostic {C}loud.
\newblock In {\em Proceedings of the 20th {ACM} Symposium on Access Control
  Models and Technologies, Vienna, Austria, June 1-3, 2015}, pages 39--50,
  2015.

\bibitem{DBLP:conf/sacmat/SquicciariniRZ18}
A.~C. Squicciarini, S.~M. Rajtmajer, and N.~Zannone.
\newblock {M}ulti-{P}arty {A}ccess {C}ontrol: {R}equirements, {S}tate of the
  {A}rt and {O}pen {C}hallenges.
\newblock In {\em Proceedings of the 23nd {ACM} on Symposium on Access Control
  Models and Technologies, {SACMAT} 2018, Indianapolis, IN, USA, June 13-15,
  2018}, page~49, 2018.

\bibitem{DBLP:conf/sacmat/WagnerBB18}
P.~G. Wagner, P.~Birnstill, and J.~Beyerer.
\newblock {D}istributed {U}sage {C}ontrol {E}nforcement through {T}rusted
  {P}latform {M}odules and {SGX} {E}nclaves.
\newblock In {\em Proceedings of the 23nd {ACM} on Symposium on Access Control
  Models and Technologies, {SACMAT} 2018, Indianapolis, IN, USA, June 13-15,
  2018}, pages 85--91, 2018.

\bibitem{DBLP:journals/access/WangZZ18a}
S.~Wang, Y.~Zhang, and Y.~Zhang.
\newblock A blockchain-based framework for data sharing with fine-grained
  access control in decentralized storage systems.
\newblock {\em {IEEE} Access}, 6:38437--38450, 2018.

\bibitem{DBLP:journals/network/XueHMWHY18}
K.~Xue, J.~Hong, Y.~Ma, D.~S.~L. Wei, P.~Hong, and N.~Yu.
\newblock Fog-aided verifiable privacy preserving access control for
  latency-sensitive data sharing in vehicular cloud computing.
\newblock {\em {IEEE} Network}, 32(3):7--13, 2018.

\bibitem{DBLP:journals/fcsc/ZuoXQZ17}
Q.~Zuo, M.~Xie, G.~Qi, and H.~Zhu.
\newblock Tenant-based access control model for multi-tenancy and sub-tenancy
  architecture in software-as-a-service.
\newblock {\em Frontiers Comput. Sci.}, 11(3):465--484, 2017.

\end{thebibliography}
}
\end{document}